\documentclass[journal,comsoc]{IEEEtran}

\usepackage{xcolor,soul,framed} 
\usepackage[justification=centering]{caption}

\colorlet{shadecolor}{yellow}
\usepackage[pdftex]{graphicx}
\graphicspath{{../pdf/}{../jpeg/}}
\DeclareGraphicsExtensions{.pdf,.jpeg,.png}

\newcommand{\red}[1]{{\color{red}{#1}}} 

\usepackage{units}
\usepackage[cmex10]{amsmath}
\usepackage{array}
\usepackage{mdwmath}
\usepackage{mdwtab}
\usepackage{eqparbox}
\usepackage{url}
\usepackage{mathtools}
\usepackage{geometry}
\usepackage{hyperref}
\usepackage{xcolor}
\usepackage{cite}
\usepackage{multirow}
\usepackage{multicol}
\usepackage{booktabs}
\usepackage{subfig}
\usepackage[T1]{fontenc}
\usepackage{lipsum}
\usepackage[utf8]{inputenc}
\usepackage{array}
\usepackage{lettrine}
\usepackage{authblk}
\usepackage{graphicx}
\usepackage{stfloats}
\usepackage{amsmath}
\usepackage{epsfig}
\usepackage{amssymb}
\usepackage{caption}
\usepackage{comment}
\usepackage[utf8]{inputenc}
\usepackage[mathletters]{ucs}
\usepackage{tablefootnote}

\usepackage[skip=2pt,font=footnotesize]{caption}

\begin{document}
   \title{Terahertz-Band Non-Orthogonal Multiple Access: System- and Link-Level Considerations
   }
    	\newgeometry {top=25.4mm,left=19.1mm, right= 19.1mm,bottom =19.1mm}%
\author{Ahmed Magbool,~\IEEEmembership{Student Member,~IEEE,}
        Hadi~Sarieddeen,~\IEEEmembership{Member,~IEEE,}
        Nour Kouzayha,~\IEEEmembership{Member,~IEEE,}
        Mohamed-Slim Alouini,~\IEEEmembership{Fellow,~IEEE, }
        and Tareq~Y.~Al-Naffouri,~\IEEEmembership{Senior Member,~IEEE,} %
\thanks{The authors are with the Computer, Electrical and Mathematical Sciences and Engineering Division (CEMSE), King Abdullah University of Science and Technology (KAUST), Thuwal 23955-6900, Makkah Province, Saudi Arabia. E-mail:\{ahmed.magbool, hadi.sarieddeen, nour.kouzayha, slim.alouini, tareq.alnaffouri \}$@$ kaust.edu.sa}}
\maketitle

\begin{abstract}
 Non-orthogonal multiple access (NOMA) communications promise high spectral efficiency and massive connectivity, serving multiple users over the same time-frequency-code resources. Higher data rates and massive connectivity are also achieved by leveraging wider bandwidths at higher frequencies, especially in the terahertz (THz) band. This work investigates the prospects and challenges of combining these algorithmic and spectrum enablers in THz-band NOMA communications. We consider power-domain NOMA coupled with successive interference cancellation at the receiver, focusing on multiple-input multiple-output (MIMO) systems as antenna arrays are crucial for THz communications. On the system level, we study the scalability of THz-NOMA beamforming, clustering, and spectrum/power allocation algorithms and motivate stochastic geometry techniques for performance analysis and system modeling. On the link level, we highlight the challenges in channel estimation and data detection and the constraints on computational complexity. We further illustrate future research directions. When properly configured and given sufficient densification, THz-band NOMA communications can significantly improve the performance and capacity of future wireless networks.
\end{abstract}

\IEEEpeerreviewmaketitle
\section{Introduction} \label{sec:intro}


Following the successful deployment of millimeter-wave (mmWave) communications in fifth-generation (5G) wireless systems, the terahertz-(THz) band is being investigated as an enabler that supports larger bandwidths between $0.3$ and $\unit[10]{THz}$ in 6G and beyond. However, in addition to the gap in THz device technologies, several THz signal processing challenges should be first overcome \cite{16}. THz signal propagation results in narrow beams, high path losses, low diffraction, high scattering, high sensitivity to blockages, and significant differences between line-of-sight (LoS) and non-line-of-sigh (NLoS) path gains \cite{ch.model}. Infrastructure enablers such as multiple-input-multiple-output (MIMO) technology can mitigate such problems by packing many antennas in arrays of tiny footprints at the transmitter and the receiver, enhancing the beamforming and multiplexing gains despite the ill-conditioned channels (see \cite{16} and references therein).

Orthogonal multiple access (OMA) techniques have been traditionally utilized to enhance system throughput and satisfy the required quality of service (QoS) requirements. OMA systems serve multiple users using orthogonal time, frequency, or code resources to prevent intra-cell interference between users. However, such orthogonality limits the number of users served simultaneously by the same base station (BS). Non-orthogonal multiple access (NOMA) techniques have been proposed to enhance spectral efficiency, distinguishing users in the power domain. In NOMA, multiple users are allowed to share the same time-frequency-code resources, and interference is mitigated using signal processing techniques \cite{50,26}. Although spectrum scarcity is not the primary concern at the THz band, there are several motivations for THz-NOMA. Due to hardware limitations, the THz band has not been fully conquered yet, and hence efficient utilization of the available bandwidths remains crucial. Higher spectral efficiency sustains higher data rates using fewer costly THz resources. Moreover, diversity in the power domain could help NOMA users to overcome the highly correlated THz channels of users. Furthermore, THz-NOMA can enhance user fairness by bridging the severe power difference between the received signals of users at different distances from the BS through proper power allocation schemes \cite{26}.

THz-band NOMA is challenging because of the narrow THz beamwidths, high-dimensional ill-conditioned (correlated) THz channels, and processing complexity (mainly in detection) under Tbps constraints. The main challenges and their prospective solutions are detailed in Table \ref{table:1}. In this article, we discuss such challenges, starting by describing the THz-NOMA system model in Sec.~\ref{sec:SysMdl}. We address system-level considerations in Sec.~\ref{sec:SysLvl}, including beamforming, user clustering, spectrum and power allocation, and using stochastic geometry as a performance analysis tool. Afterward, we discuss link-level considerations of channel estimation and data detection in Sec.~\ref{sec:LnkLvl}. We provide end-to-end simulations of THz-NOMA systems in Sec.~\ref{sec:sim}, demonstrating the performance and complexity tradeoffs. We then compare NOMA with another broadcasting scheme, namely multi-user linear precoding in Sec.~\ref{sec:MU-LP}. We illustrate future research directions and conclude the paper in Sec.~\ref{sec:con}.

\begin{table*}
\footnotesize
\centering
\caption{Main THz-band NOMA challenges and mitigation strategies (Numbers from TeraMIMO channel simulator \cite{ch.model}; carrier frequency of $\unit[300]{GHz}$).}
\begin{tabular} {||m{3cm}|m{6cm} | m{5cm} ||}
 \hline Challenge & Example & Mitigation strategies / opportunities \\  [0.5ex]
 \hline
 Narrow THz beamwidths &  For a $\unit[25]{dBi}$ sub-array gain, the azimuth and elevation angular spreads do not exceed $11.42^{\circ}$ \cite{ch.model} & Beamwidth control - multi-beamforming \\
 \hline User fairness & A user at $\unit[5]{m}$ distance from the transmitter experiences  \unit[14]{dB} more path loss than another user at \unit[1]{m} (according to the free space model) &  Proper user clustering and power and spectrum allocation\\
 \hline
 High-dimensional channels & It is typical to transmit 64 data streams over 64 SAs at both the transmitter and the receiver, with 16 antenna elements per SA to achieve a channel gain of $0.0163$ for a user at $\unit[5]{m}$ & Computationally efficient system-level and link-level signal processing algorithms\\
 \hline
 Low-rank channels &  Channels in the true THz range are of rank $\sim$ 1 & Leveraging sparsity in the spatial domain to reduce the complexity of baseband algorithms (fast channel updates; exploiting the spherical wave model)\\
 \hline
  High channel correlation across users & Users at distances of \unit[1]{m} and \unit[5]{m} form the transmitter and $20^{\circ}$ elevation and angular spreads have channels with a correlation coefficient of 0.97 & Reducing baseband complexity by eliminating redundant computations across users  \\
 \hline
 Terabit per second (Tbps) detection &  At existing circuit clock speeds, $\sim$1000 bits need to be processed per clock cycle to achieve a Tbps \cite{26} & Computationally efficient SIC data detectors and decoders \\
 \hline
\end{tabular}
\label{table:1}
\end{table*}

\section{System Model} \label{sec:SysMdl}

Although MIMO architectures are used in lower frequencies to enhance the beamforming and multiplexing gains, their use in THz systems is more crucial to compensate for the severe path loss. Both the transmitter and the receiver in a THz system employ a large number of antennas, typically organized in an array-of-subarray (AoSA) architecture; each subarray (SA) transmits one information symbol, and multiplexing is enabled over different SAs \cite{16}. Thanks to the tiny wavelengths of THz waves, such multiple-antenna designs can be realized in small device footprints.

\begin{figure*}
         \centering
         \includegraphics[width=.95\textwidth]{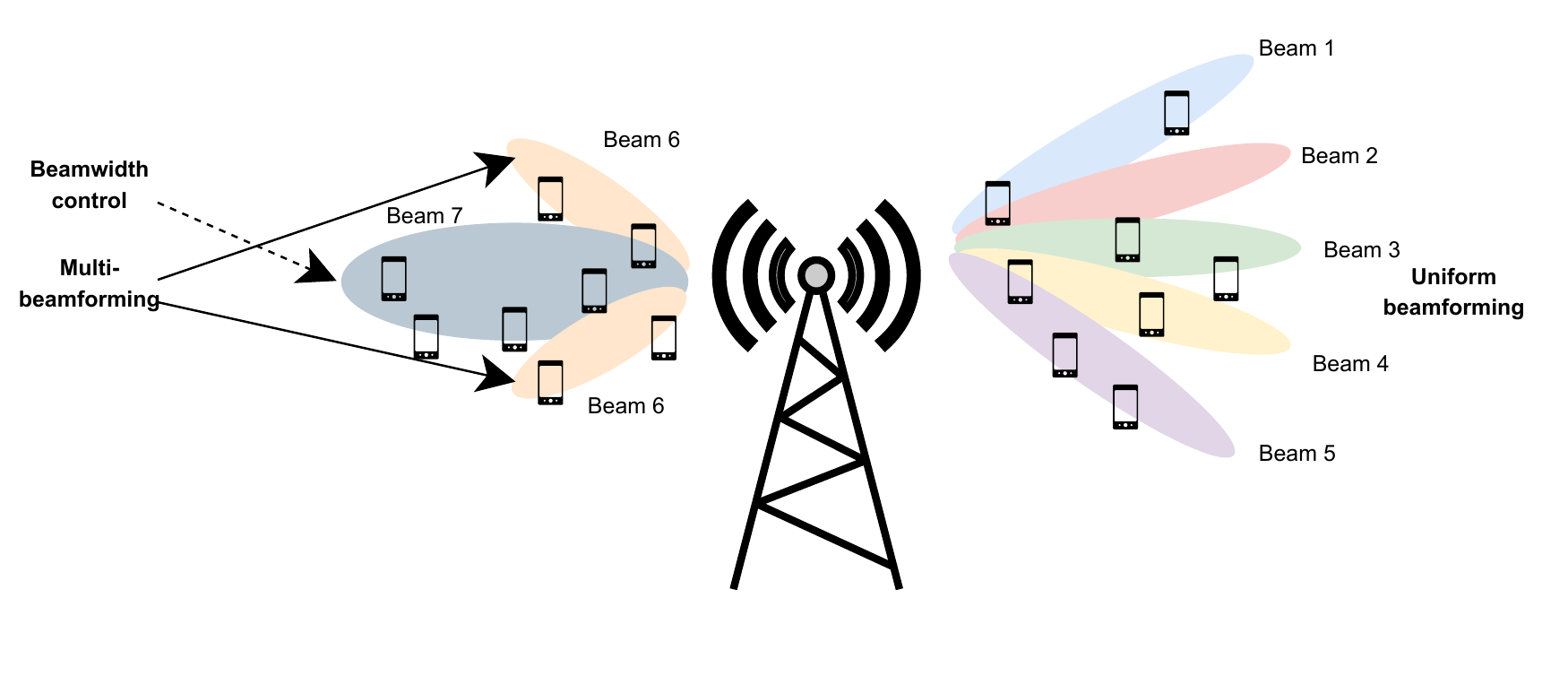}
        \caption{NOMA systems with uniform beamforming, beamwidth control, and multi-beamforming.}
        \label{fig:NOMA}
        \vspace{-0.2cm}
\end{figure*}

The first step of THz-NOMA is clustering users into groups that share communication resources, which is very challenging given the narrow width of the THz beams. Each user cluster is typically served by a single THz beam in beam-division multiple access (BDMA), as illustrated in Fig.~\ref{fig:NOMA}~\cite{50}. However, only a small number of users can be served within a single NOMA group because user interference severely degrades the quality of decoded signals and causes a non-negligible delay in successive interference cancellation (SIC). Alternatively, users per beam can be further divided into sub-clusters (groups), where orthogonal group resources avoid inter-group interference (i.e., OMA between groups and NOMA within a group).

The second step of NOMA is power allocation. For simplicity of illustration, we consider two-user NOMA groups served by a BS over the same time-frequency-code resources. Superposition coding combines the users' signals before transmission, assigning a higher power level to the user with worse channel conditions. Because THz channels are LoS-dominant and subject to a severe path loss, the distance between users and the BS dictates to a great extent the channel condition.


The third step of NOMA is successive interference cancellation (SIC) at the receiver side. Since the weak user (of worse channel conditions) is assigned higher power, its component in the received signal is more significant; it can directly decode its signal, treating the other user's signal as noise. The strong user, on the other hand, first decodes the weak user's signal, treating its own signal as noise, and then applies SIC to remove the interfering signal and decode its own signal.

The benefits of THz-NOMA can only be fully realized following optimizations and efficient signal processing at both the system and link levels. We address the corresponding system- and link-level prospects and challenges in Sections \ref{sec:system} and \ref{sec:link}, respectively.

\section{System-Level Considerations} \label{sec:SysLvl}
\label{sec:system}

At the system level, we consider multi-user beamforming, NOMA user clustering, spectrum and power allocation, and motivate the use of stochastic geometry as a performance analysis technique.

\subsection{Beamforming}



\begin{figure}
         \centering
         \includegraphics[width=.48\textwidth]{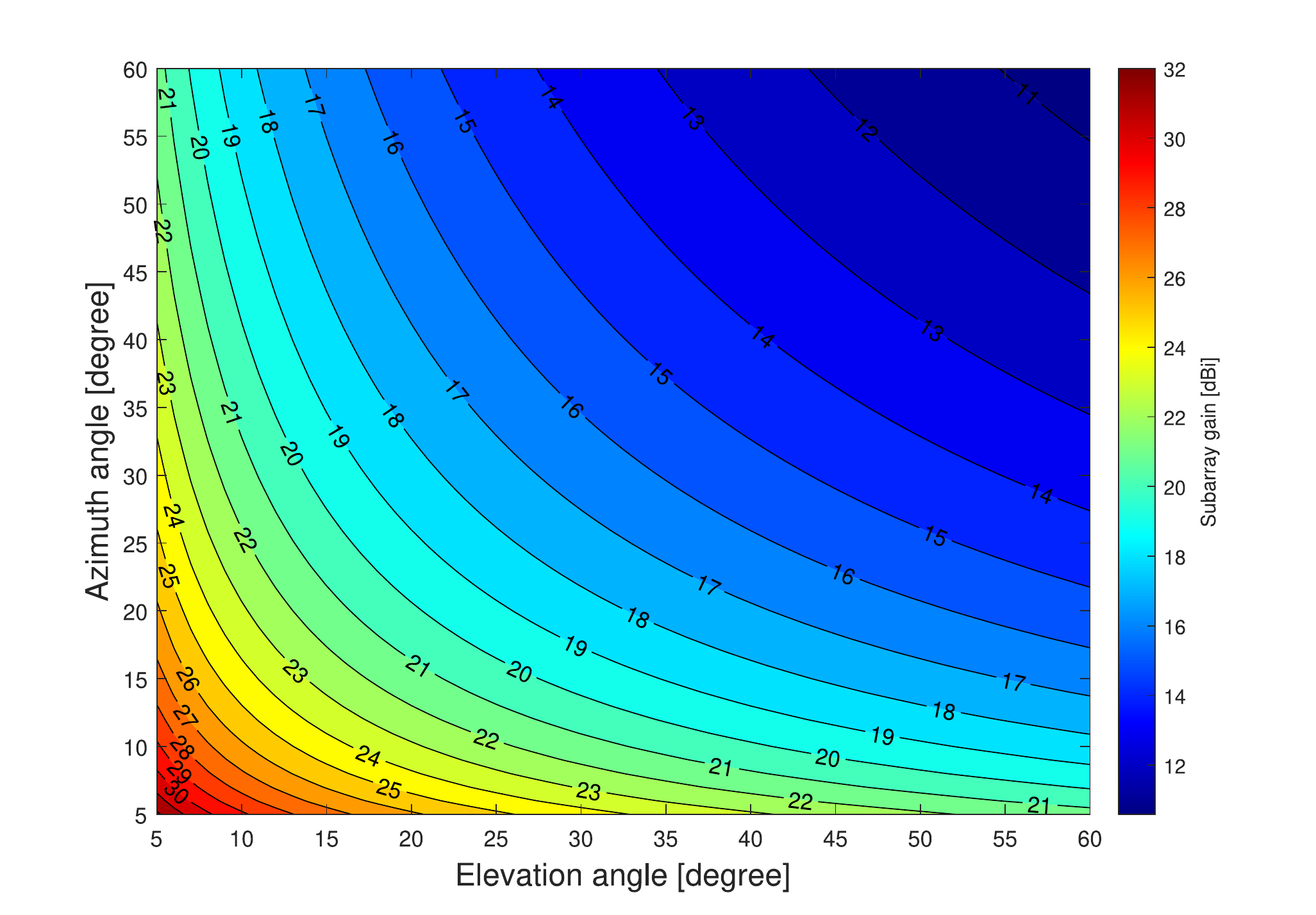}
        \caption{Heat map of the subarray gain in dBi as a function of the azimuth and elevation angular spreads.}
        \label{fig:AoD}
\end{figure}

THz-NOMA beamforming aims at forming multiple beams of proper direction and width to serve multiple users. The half-power elevation and azimuth beam angles are inversely proportional to the SA beamforming gain \cite{ch.model}, as illustrated in Fig.~\ref{fig:AoD}. For example, a beam with elevation and azimuth angular spreads of $27^{\circ}$ can have a SA gain of $\unit[17.3]{dBi}$, which is usually not sufficient for medium-distance communications~\cite{ch.model}. The probability of serving $k$ uniformly distributed users (polar uniform distribution) in an angular spread of $27^{\circ}$ is small ($0.02$ for $k\!=\!2$; so an average of $100$ users are required to serve two users per beam). Such observations raise concerns about the practicality of NOMA at high-frequency bands.

Beamwidth control techniques~\cite{104} can increase the probability of serving multiple users by widening the main lobe (e.g., beam $7$ in Fig.~\ref{fig:NOMA}), for example. THz-band ultra-massive MIMO (UM-MIMO) beamwidth control should achieve a trade-off between complexity and performance. Simple solutions such as conventional beamforming (CBF), where the number of active antennas is reduced to widen the analog beamwidth with constant modulus phased sifters (PSs), are favorable due to their low complexity. However, such schemes lack the flexibility to change beamwidths with user activities and densities. More computationally complex optimization-based beamwidth control methods can maximize the worst-user system sum rate (max-min fairness (MMF)).
Furthermore, multi-beamforming methods can steer a beam generated from one radio-frequency (RF) chain to serve angularly distant users (e.g., beam $6$ in Fig.~\ref{fig:NOMA})~\cite{122}. However, multi-beamforming reduces the SA gain as the transmit power will be split into different physical directions.

There are a number of THz-NOMA beamforming solutions in the literature. Simple approaches such as uniform beamforming minimize system energy consumption but are not efficient solutions at low user densities as energy can be wasted forming beams in no-user directions. A more advanced solution in~\cite{NOMA_BF1} first clusters users and then forms beams with appropriate directions and widths that suit the NOMA groups. In particular, the beamforming problem is jointly optimized with power and bandwidth allocation following user clustering to maximize system throughput.

\subsection{User Clustering} \label{sec:UC}
Clustering is the task of distributing users into small groups in which NOMA is applied. Although BDMA mitigates inter-beam interference, the interference between different groups within the same beam should still be avoided, perhaps using OMA techniques. A basic user-pairing approach is based on the corresponding user channel conditions, where weak users are typically paired with strong ones. If the LoS is blocked, the corresponding user is directly classified as weak as the difference between THz LoS and NLoS components is significant. Otherwise, users are classified based on their distances from the BS. So users inside a circle of radius $r$ (smaller than cell radius) are strong users and the remaining ones are considered weak. Distance-based user distinction is particularly plausible at THz frequencies due to severe path losses.

Similar-distance grouping (SDG)~\cite{NOMA_BF1} is another candidate scheme for THz-NOMA clustering. In SDG, users are clustered according to their distances from the BS; the nearest two users form the first group, the second nearest the second group, etc. As opposed to strong-weak user pairing, SDG pairs a strong user with another strong user and vice versa. Consequently, spectrum allocation can be facilitated, where weak-user groups get assigned sub-bands suffering from less path loss (more details in the next subsection).

Clustering can be updated rather than performed from scratch every time by utilizing proper clustering update algorithms. By doing so, an optimization-based clustering technique can be solved to determine the initial association of the users and then a clustering update technique can be invoked based on the users' mobility without the need to solve the optimization problem again. Unsupervised and supervised machine-learning and reinforcement learning tools can also be used to cluster users based on features such as geographical location and channel condition to indicate the best user clustering policy.

\subsection{Spectrum Allocation}

Spectrum allocation is the problem of assigning carrier frequencies to NOMA groups. Frequency-based OMA techniques such as orthogonal frequency-division multiple access (OFDMA) can still be used to distinguish NOMA groups within a beam. However, due to THz-band high frequency-selectivity, hardware constraints and molecular absorption, THz path losses are more severe at particular sub-bands that need to be avoided. The available bandwidth is thus divided into smaller transmission windows, the centers of which enjoy less path loss than the sides~\cite{1004}. For instance, a total bandwidth of $\unit[45]{GHz}$ is available for transmission at a central frequency of $\unit[300]{GHz}$ (transmission window from $227.5$ to $\unit[322.5]{GHz}$)~\cite{1004}. Efficient THz carrier frequency assignment in OMA can leverage this property to allocate window centers to faraway users and window sides to closer users in a distance-aware multi-carrier (DAMC) approach~\cite{1004}. The same concept can be applied to NOMA by allocating each NOMA group a sub-band at the transmission window center or side, depending on the distance from the BS. To this end, a group distance metric needs to be defined, the center of mass of user distances in a group, for example. The distinction between the group distances is more clear if users are clustered using the SDG scheme described in Sec. \ref{sec:UC}.

\subsection{Power Allocation} \label{ssec:PA}
Power allocation is the process of determining the power to be assigned for each NOMA user, with the objective of maximizing sum rate, energy efficiency, or system fairness. In an AoSA architecture, power allocation can be expressed in terms of power assigned per transmitting SA, accounting for antenna and beamforming gains (per SA). Conventional power allocation techniques are constant or optimization-based. Intuitively, users with worse channel conditions should be assigned higher power levels for fairness, and low complexity assignments (without the need for solving complex optimizations) are favored. For example, users can be ordered based on their channel conditions; the user with the worst condition gets assigned a certain power, the second weak user a fraction of that power; and continue this way while observing the maximum transmit power constraint~\cite{14}. In optimization-based solutions, power allocation can be coupled with beamforming and user pairing. However, solving such global optimizations is computationally complex, especially using iterative algorithms. Besides, such optimizations are usually non-convex, and a global solution is not guaranteed.

As an illustration, consider a NOMA group with three users at distances $1$, $1.2$, and $\unit[7]{m}$ from the BS; assume a transmit power budget of $\unit[20]{mW}$ and a power fraction of $0.5$. The power allocated to the furthest, middle, and nearest user is $11.43$, $5.71$, and $\unit[2.86]{W}$, respectively, which results in a receive power of $-88$, $-76$ and $\unit[-77.4]{dBm}$ according to the free space model. This approach clearly lacks fairness, which can be enhanced by linking power allocation to user channel gains, as in fractional transmit power allocation (FTPA)~\cite{14}, for example. Applying FTPA to our example, with the assumption that the channel gain is inversely proportional to the square of the distance from a BS, the power distribution becomes $19$, $0.6$, and $\unit[0.4]{mW}$ for the furthest, middle, and nearest user, respectively. This will result in a received power of around $\unit[-86]{dBm}$ for all users.



\subsection{Stochastic Geometry for THz-NOMA}



Stochastic geometry (SG) can be used as a powerful model-driven tool that provides a mathematical framework to capture the properties of THz networks and quantify the benefits of THz-NOMA. Several research works have leveraged SG for the performance analysis of power-domain NOMA networks operating on sub-$6$~GHz and mmWave frequency bands~\cite{SG_NT}. However, there is still a gap in the literature in extending the available models to account for specific characteristics of THz-NOMA. In the SG analysis of multi-cell NOMA systems, clustering algorithms that capture the users to be simultaneously served in the same resource block should be clearly defined. This is important because the distances between the users and the serving BS within a NOMA group are dependent on the adopted clustering technique \cite{79}. A more general setup is also considered by modeling the users as a Poisson cluster process (PCP). The beamforming technique and the narrow beamwidth will affect user clustering and hence the distance distributions of the serving and the interfering BSs. The accuracy of the SIC process at the receiver should also be taken into consideration, which can be modeled by introducing a fraction between $0$ and $1$ in the SINR expression (for more details, see \cite{SG_NT} and the references therein).

In addition to previous modeling choices, several modifications should be applied to conventional SG frameworks to account for THz-specific characteristics in THz-NOMA networks. Typically: 1) THz antennas should be modeled as directional antennas. 2) 3D modeling of THz networks is mandatory as it accounts for the effect of vertical heights. While it may be acceptable to ignore the heights in sub-$6$~GHz networks as they enjoy large communication distances, such an assumption is inaccurate for dense THz networks with limited communication ranges. This will require the use of 3D directional antenna models such as the 3D pyramidal-plus-sphere-sectored antenna model~\cite{SG_NT,SG2} at THz systems. 3) The impact of blockages can be captured by a LoS/NLoS probabilistic model that identifies each link's condition based on the link length and the propagation environment. In THz systems, blocking LoS paths is mainly due to attributes such as dynamic self-blockage and human blockage and static blockages like buildings in outdoor environments and walls or furniture in indoor environments~\cite{SG_NT,SG2} 4) Since THz beams are usually narrow and as THz communication is more sensitive to blockages compared to sub-$6$~GHz, the small scale fading can be ignored in the analysis of THz networks. 5) As opposed to lower frequencies, THz communication is affected by water molecules in the environment. To account for this, the THz propagation model is a frequency-dependent exponential function that highlights the effect of both molecular absorption loss and spreading loss \cite{SG_NT,SG2}. 6) Finally, SG needs to account for thermal noise as well as molecular absorption noise caused by the re-radiation of the absorbed signal energy.

\section{Link-Level Considerations} \label{sec:LnkLvl}
\label{sec:link}
At the link level, we need to consider the problems of efficient channel estimation and data detection.

\subsection{Channel Estimation}
Channel state information (CSI) is required at the transmitter for beamforming, user pairing, and power allocation, and at the receiver for SIC and symbol decoding. Imperfect channel estimation causes power allocation ambiguity which reduces the NOMA sum rate. It is essential to investigate the effect of channel mismatch caused by THz propagation characteristics such as beam squint and spherical wave propagation. Due to the large number of antennas at both the transmitter and receiver, computationally efficient channel estimation solutions are crucial \cite{ch.model}.

THz signal propagation models should also account for some widely ignored characteristics at lower frequency bands. For instance, spherical wave propagation  (SWP) models should replace plane wave approximations that are inaccurate at THz frequencies and may lead to channel estimation errors in massive MIMO systems. In particular, hybrid spherical- and planar-wave channel modeling solutions are found to be sufficiently accurate and less complex than fully spherical wave models \cite{41}. Similarly, the beam squint effect, caused by frequency-independent steering vectors in wide-band massive MIMO THz beamforming, should be accounted for \cite{44}; ignoring such frequency dependency results in signals being steered into unintended physical directions.

When accounted for, beam squint and SWP can result in enhanced THz-NOMA system performance. For instance, the beam squint effect can be exploited to serve more users as beams are split in different directions (a concept similar to multi-beamforming). Moreover, the SWP model can better distinguish the channels of different users as it reduces the correlation between these channels (i.e., SWP introduces more degrees of freedom). Such characteristics are useful when modeled carefully and integrated into the system model.

In addition, despite being problematic for NOMA, the high correlation between user channels can be leveraged to reduce channel estimation complexity significantly. For example, instead of estimating the CSI for every user independently, minimal channel estimation updates on a specific CSI instance based on geometry information can construct the channels of the neighboring users. The computational complexity of the channel estimation process can be reduced using multiple methods. In general, the low-rank LoS-dominated THz channels can convert the channel estimation problem into a sparse recovery problem in the angle domain. Since THz channels have usually low ranks, the single-input-single-output (SISO) channel can be estimated between one transmitting and one receiving SA and appropriate tuning is performed to find the other entries of the channel matrix. Furthermore, channel update algorithms can be utilized to expand the channel estimated in one spatial, temporal or spectral domain to others.

\subsection{Data Detection}
While channel coding and decoding account for most of the computational cost in the THz baseband, a proper design of data detectors can significantly reduce the complexity of THz UM-MIMO systems. When deciding on the detector of choice, it is important to note the constraints on the processing capabilities towards achieving the promised Tbps data rates. With a few GHz clock speeds, baseband processing should be executed at a rate of at least 1000 bits per clock cycle to achieve Tbps. Existing THz transceivers are not yet capable of supporting such rates. On top of detection constraints in conventional THz systems, SIC in NOMA at least doubles (in two-user NOMA) the resources required to decode signals; strong users have to detect both their symbols and those of the weak users, all while satisfying the Tbps constraint of THz communications. Due to the corresponding SIC delay, having more than two users in a THz-NOMA group is impractical.

The maximum-likelihood (ML) detector is optimal but is incapable of meeting the Tbps constraint because of its computational complexity (exponential increase with MIMO dimension). Furthermore, unlike conventional low-frequency massive MIMO systems, the high-dimensional (doubly-massive) ill-conditioned THz channels make linear detectors such as zero-forcing inefficient (suffering from performance loss caused by inverting low-rank channel matrices). THz MIMO systems thus require more sophisticated, yet low-complexity non-linear detectors such as nulling-and-cancellation (NC), chase detectors, and the layered orthogonal lattice detector (LORD), all of which are realizations of subspace detectors \cite{26}. These detectors can be employed to puncture the channel and force certain structures to help in the data detection. In particular, subspace decompositions can provide sufficient parallelizability in the detector and SIC NOMA stages to achieve the Tbps complexity and memory requirements \cite{26}. In near-static THz scenarios where the channel is flat, preprocessing matrix decompositions can be retained across multiple frames, further reducing complexity. Moreover, even under slight channel variations, matrix decompositions can be updated without recomputing them from scratch.

\section{Simulations and Discussion} \label{sec:sim}
In this section, we provide end-to-end simulations of THz-NOMA systems addressing system-level and link-level considerations.
\subsection{System-Level Simulation}
We consider $n$ users randomly and uniformly distributed inside a 2-D circular cell of radius $\unit[10]{m}$. The SA beamforming gain is set to $16.61 \unit{dBi}$, which results in beams with half-power azimuth and elevation angles of $30 ^{\circ}$. The transmission window is centered at $\unit[300]{GHz}$ with a total bandwidth of $\unit[10]{GHz}$. We adopt uniform beamforming, so $12$ beams are needed to cover the cell.

At the system level, we compare three THz-NOMA baselines: (1) distance-based clustering and power allocation \cite{26}, (2) exhaustive-search-based user clustering and power allocation that maximizes user fairness per beam, and (3) THz-OMA with frequency-division-multiple-access (FDMA). Each NOMA group is assigned $1$ GHz; each OMA user is allocated $1$ GHz. The channels are generated using the TeraMIMO channel simulator \cite{ch.model}; four transmitting and receiving SAs; $64$ antennas per SA; carrier frequency of $\unit[300]{GHz}$. Although the channel dimension is $4 \times 4$, this is still considered an UM-MIMO system as the transmitter and the receiver consist of $256$ antenna elements each. We assume $20$ or 50 users per cell. Since $12$ beams are required to cover the cell, on average, $24$ users should exist in a cell for allocating two users per NOMA beam.
 \begin{figure}
         \centering
         \includegraphics[width=.48\textwidth]{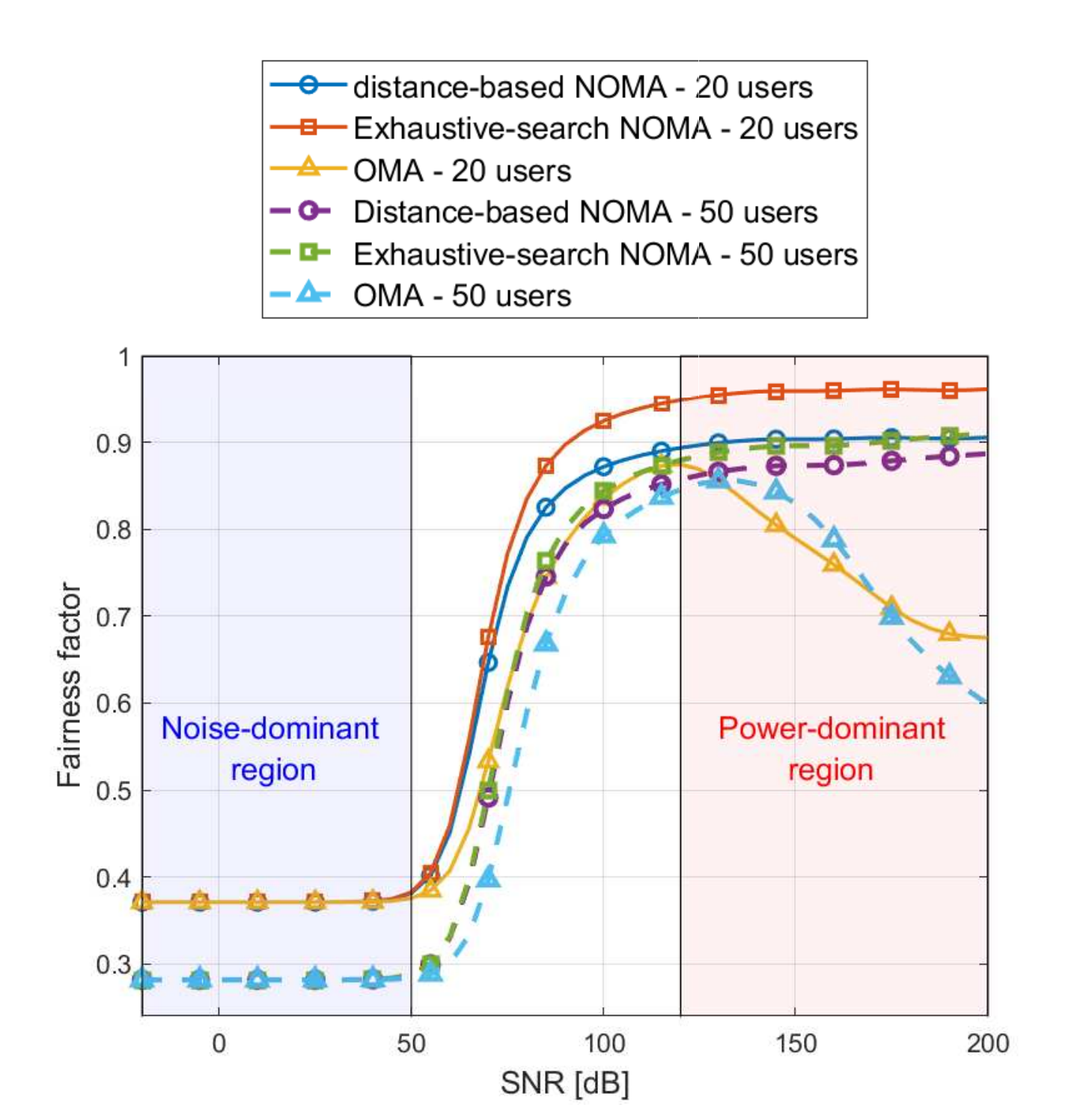}
        \caption{Fairness factor for NOMA and OMA systems with 20 and 50 users.}
        \label{fig:FF}
\end{figure}

Given that spectral efficiency is not the main bottleneck of THz systems due to the huge bandwidth availability, we examine user fairness with NOMA. We introduce a fairness metric, computed as the ratio of the lowest and highest rates in the beam. The fairness metric is 1 for equal maximum and minimum user rates (maximum fairness) and 0 for a lowest user rate of zero; a fairness metric closer to $1$ indicates even distribution of user rates. A weighted average by the number of users per beam gives the overall fairness factor of the system.

Fig. \ref{fig:FF} depicts the fairness versus the signal-to-noise ratio (SNR) defined as the ratio of total transmit power by the noise power. The fairness factor increases with the number of users, which is intuitive and further highlights the importance of user densification in THz systems. Furthermore, distance-based NOMA outperforms OMA, and exhaustive-search-based NOMA forms an upper bound on NOMA system fairness. The results can be partitioned into three regions based on SNR. The first region (lower SNR regime) is noise-limited and does not achieve good fairness. The second region introduces fairness improvements to all schemes. However, in the third region (high SNR regime), OMA fairness starts to decay. Therefore, despite the slight increase in the fairness factor of NOMA systems at high SNR, NOMA algorithms can still maintain user fairness.




\subsection{Link-Level Simulation}

\begin{figure}
         \centering
         \includegraphics[width=.48\textwidth]{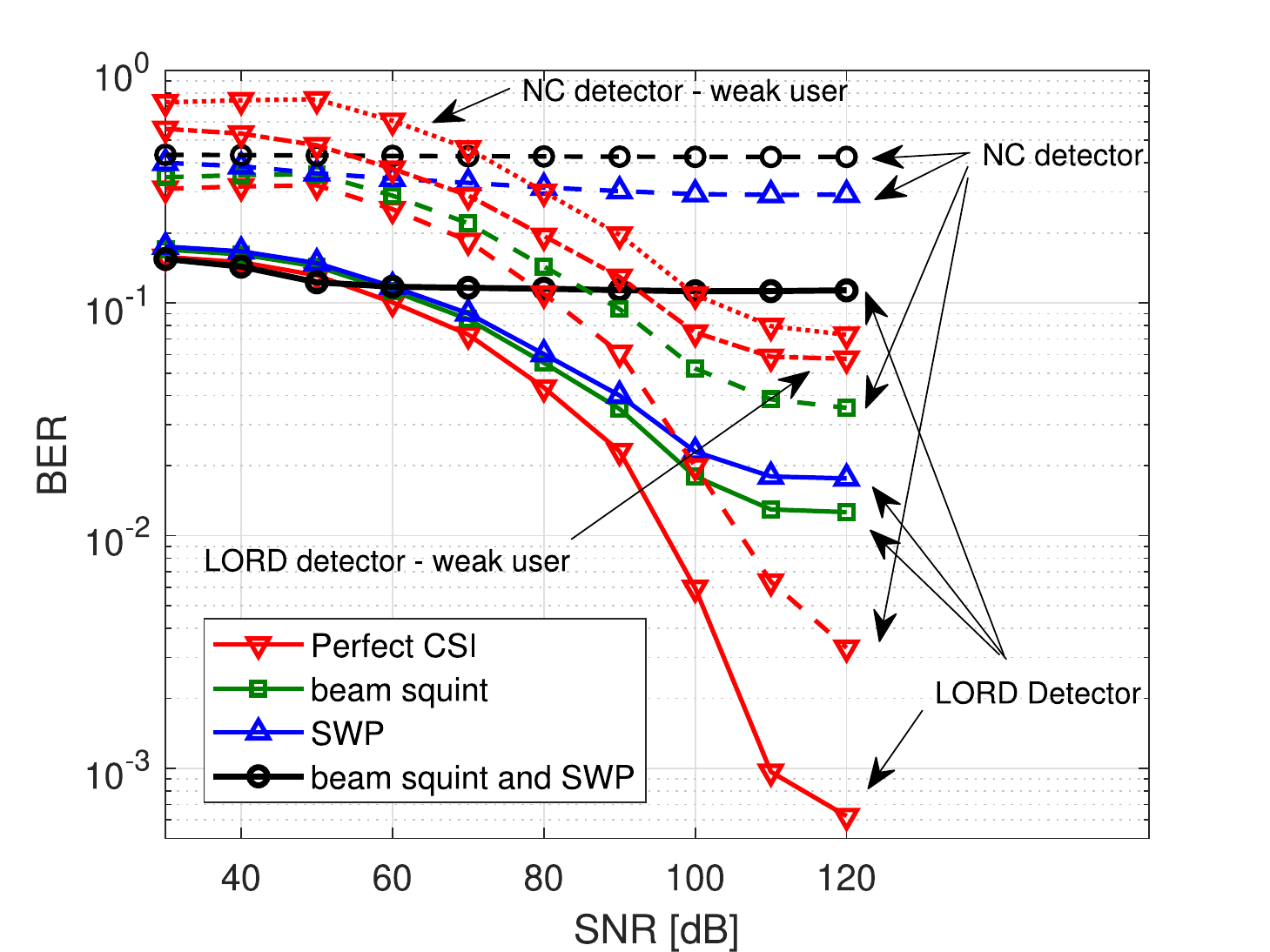}
        \caption{BER versus SNR when the detectors have full knowledge of CSI (red), do not account for beam squint (green), do not account for SWP (blue) and do not account for both beam squint and SWP (black) for LORD detector (solid) and NC detector (dashed). All curves are plotted for the strong user, whereas, for clarity, only the case of perfect CSI is illustrated for the weak user (doted-dashed for LORD detector and dotted for NC detector).}
        \label{fig:BER}
\end{figure}

For link-level simulations, we consider a cell of $\unit[10]{m}$ radius and 100 users. We compare two detectors, nulling and cancelling (NC) and LORD, assuming distance-based clustering and power allocation (see \cite{26} for more details). We study the corresponding bit-error-rate (BER) performance in Fig. \ref{fig:BER} assuming"LoS + multi-path" THz channels, and accounting for CSI errors caused by ignoring beam squint and SWP. LORD generally outperforms NC at a marginal complexity cost. It can be noted that the weak user has a significantly higher BER compared to the strong user. This can be attributed to two main reasons. First, as opposed to the strong user, the weak user does not apply any form of interference cancellation. Second, the weak user has a much lower channel gain than the strong one due to the severe path loss. The error of detecting the bits for the weak user can be propagated to degrade the performance of the BER of the strong user as well. Furthermore, the system performance is degraded by ignoring beam squint, more significantly and more by ignoring SWP. The BER can increase by more than an order of magnitude when both beam squint and SWP are neglected compared to the case of perfect CSI. This observation further highlights the importance of designing joint THz-NOMA clustering, power allocation, channel estimation, and detection algorithms.

\section{NOMA versus MU-LP} \label{sec:MU-LP}

The achievable gains of MIMO NOMA at lower frequencies are disputed, citing equivalent or superior performance provided by the much simpler multi-user linear-precoding (MU-LP) scheme \cite{40}. MU-LP is a special class of broadcasting techniques that, unlike NOMA, do not employ SIC at any receiver. That is, a superposition signal is transmitted, and each user recovers its own data by treating other users signals' as noise.
The multiplexing gain, which measures how fast the rate changes with SNR (system performance at high SNR), is studied for NOMA and MU-LP in \cite{40}. NOMA achieves lower multiplexing gains than MU-LP under full-rank channels due to the SIC rate constraint in the latter. In particular, the NOMA transmission rate is restricted to support both paired users while preventing SIC error, which does not apply to MU-LP. This argument assumes a zero-forcing beamformer (ZFBF) to mitigate intra-cluster interference.

\begin{figure}
         \centering
         \includegraphics[width=.48\textwidth]{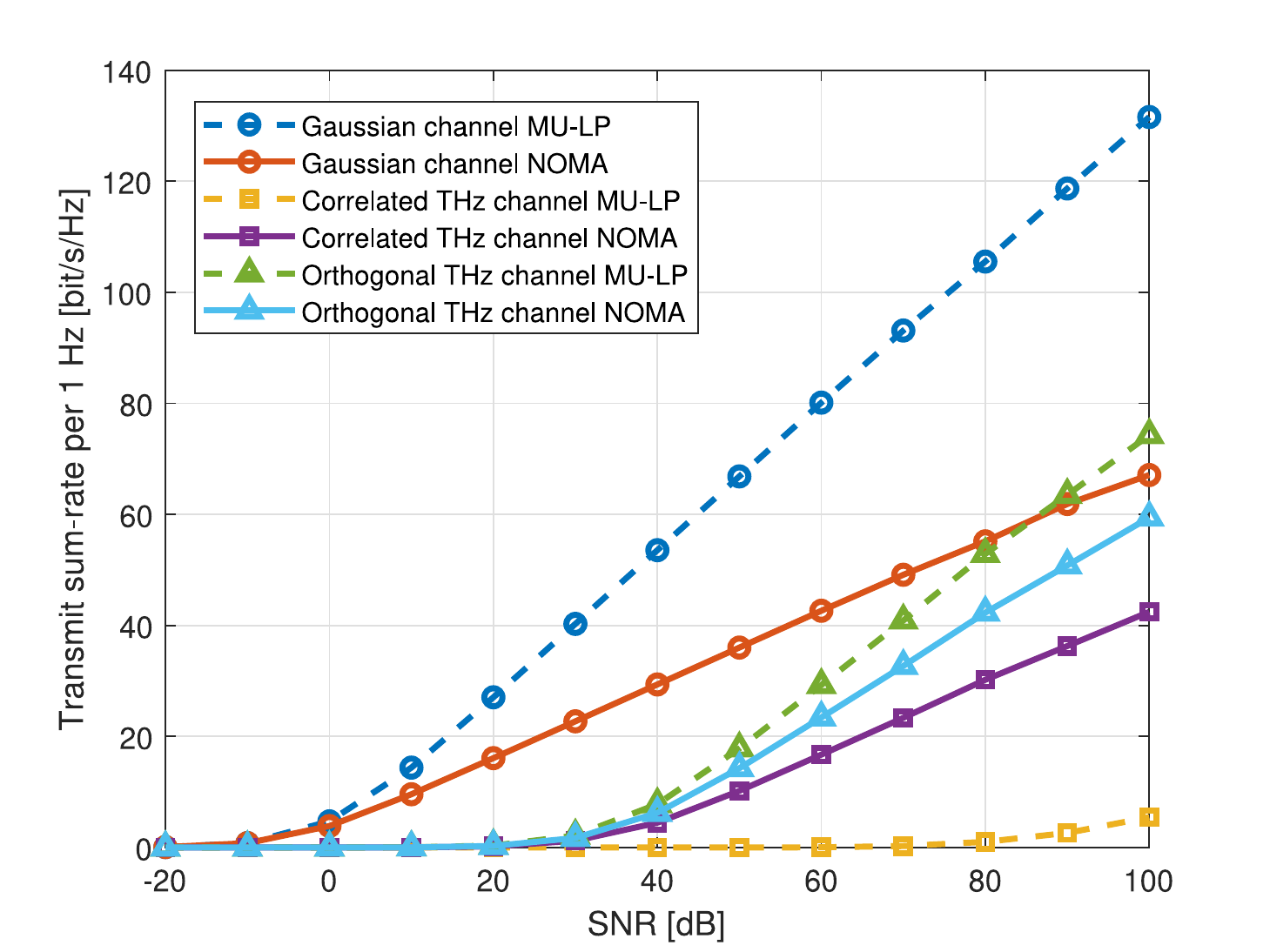}
        \caption{Sum-rate per Hz versus SNR using ZFBF-based power allocation for Gaussian and THz channels (correlated and orthogonal) for NOMA and MU-LP systems two-user systems.}
        \label{fig:ZFBF}
\end{figure}

To check the validity of this conclusion in THz scenarios, the system capacity versus SNR is simulated in Fig. \ref{fig:ZFBF} for three channels types: Gaussian (zero mean and unit variance; typical for lower frequencies), correlated, and orthogonal THz channels ($4\!\times\!4$ SA-channel at $\unit[300]{GHz}$ using TeraMIMO \cite{ch.model}). THz channels can be made orthogonal by careful spatial tuning of SA positions as described in \cite{26}. Due to high path loss, inter-cell interference has a negligible effect in the THz band compared to the lower frequencies and can be safely ignored. For MU-LP, we apply ZFBF to mitigate intra-cluster interference in both users. ZFBF is accompanied with rate-maximization to obtain the strong-user power matrix in NOMA (no interference following SIC). Fig.~\ref{fig:ZFBF} verifies that MU-LP achieves a higher multiplexing gain (slope of sum-rate) in Gaussian and orthogonal THz channels. However, this observation does not apply to naturally occurring THz correlated channels. Applying ZFBF on ill-conditioned channels to cancel the interfering-user signal corrupts the main signal of the intended user, reducing the achievable rate. Therefore, ZFBF-based MU-LP is not a valid solution at the THz-band, despite its favorable low complexity.

\section{Conclusions} \label{sec:con}

In this work, we discuss the prospects and challenges of applying NOMA techniques to THz communications. We investigate system- and link-level considerations of THz-NOMA, highlighting the effects of massive antenna arrays, high channel correlation, narrow beams, etc. We demonstrate that THz-NOMA can indeed enhance system throughput and user fairness in cells with sufficient densifications, which is a promising research direction towards 6G. Therefore, this work provides solid grounds for future THz-NOMA research that considers realistic THz-specific constraints.

Beside the future THz-NOMA research directions mentioned previously in the paper, several others can be considered to achieve ultimate network promises of maximizing area traffic capacity, spectral efficiency, user fairness, and bandwidth utility. For example, on the algorithmic level, learning-based solutions can address system-level THz constraints in beamforming and beam steering, forming beams by learning user activity and dynamicity. Furthermore, on the infrastructure level, intelligent reflecting surfaces (IRSs) can be utilized to control the propagation environment and enhance THz channel conditions. IRSs can boost the THz multiplexing gain and reduce the huge gap between the LoS and NLoS components.



\section*{Acknowledgment}

Research reported in this publication was supported by King Abdullah University of Science and Technology (KAUST).


%





\ifCLASSOPTIONcaptionsoff
  \newpage
\fi






%


\end{document}